\documentclass[twocolumn,showpacs,amsmath,amssymb,floatfix,superscriptaddress]{revtex4-1}
\usepackage{graphicx}
\usepackage{color}

\begin{document}

\title{Spaceless description of active optical media}

\date{\today}

\author{Giovanni Giacomelli}
\affiliation{Consiglio Nazionale delle Ricerche, Istituto dei Sistemi Complessi, via Madonna del Piano 10, I-50019 Sesto Fiorentino (FI), Italy}

\author{Serhiy Yanchuk}
\affiliation{Institute of Mathematics, Technische Universit\"at Berlin, Stra\ss e des 17. Juni 136, D-10623 Berlin, Germany}

\author{Antonio Politi}
\affiliation{Institute of Pure and Applied Mathematics, Department of Physics (SUPA), Old Aberdeen, Aberdeen AB24 3UE, United Kingdom}

\begin{abstract}

\noindent The acclaimed Maxwell-Bloch (or Arecchi-Bonifacio) equations are a valid dynamical model, effectively  describing
wave propagation in nonlinear optical media: from the amplification in input-output devices
to multimode instabilities arising in laser systems.
However, the inherent spatial variability of the physical observables 
represents an obstacle to fast simulations and analysis, especially
whenever networks of active elements have to be considered.
In this paper, we propose an approach which, stripping the spatial dependence of its role as a generator of dynamical richness, allows for a compelling simple portrait. It leads to (a few) ordinary differential equations in input-output configurations, complemented by a time-delayed feedback in closed-loop setups. Such scheme reproduces accurately the dynamics, paving the way to a plain treatment of the wealth of phenomena described by the Maxwell-Bloch equations.

\end{abstract}

\maketitle

\section{Introduction}

\noindent Optical active media are the pillar of a large variety of schemes and physical phenomena ranging from signal enhancement in detection setups \cite{Frede2007}, to regeneration of digital transmissions in fibers \cite{Li2017}, and chirped pulse amplification \cite{Strickland1985}. By far, the most common setup based on active media is the laser: using a cavity, the coherent amplification process combined with the feedback mechanism produces  a strong emission of radiation with striking properties. Recently, lasing networks (LANERs) have been introduced as generalization of the laser \cite{Lepri2017}, where different elements interact in a nontrivial way to determine the overall dynamical properties. At variance with standard networks often considered in the literature, LANER links have their own dynamics, and the complex character of the relevant observable (the field) is responsible for fascinating interference phenomena \cite{Giacomelli2019,Giacomelli2020}. 
The number of physical media displaying coherent optical gain is huge as well as the diversity of the underlying mechanisms (see e.g. \cite{OA1999}). Among others, of particular relevance are semiconductor amplifiers \cite{SMC} and Erbium-doped fibers \cite{ER} for their widespread applications in IT and communication infrastructures. 
Despite this variety, a unique  mathematical model capturing most of dynamical features of active media is
available since many years: the Maxwell-Bloch equations. 
Here, following Ref.~\cite{McNeil2015}, we prefer to refer to them as to the Arecchi-Bonifacio (AB) model \cite{Arecchi1965}. The AB equations are a semiclassical description, where the field is treated classically; they have been derived in the so-called slowly varying envelope approximation (i.e. after removing the optical high frequencies). 
However, the AB model involves partial differential equations; as such, it is difficult to analyze and unsuitable in simulations of long active media and in the characterization of setups composed of many elements. 

In this paper, we propose a novel approach which allows simplifying the model structure,
without losing the dynamical complexity of the original problem.
By adopting a Lagrangian viewpoint, we rewrite the AB model in a moving frame. In this representation, the spatial variation of the
various fields is a sheer, dynamically stable, amplification, described very well by low-order polynomials.
By expanding the fields into (orthogonal) Leg\`endre polynomials, the AB model is mapped onto a hierarchy of ODEs, complemented by
suitable boundary conditions.
By retaining polynomials up to order $n$, one obtains a spaceless model named SLn, involving the amplitude
of the $n$ leading Leg\`endre modes of polarization and population.

A preliminary application to an input/output setup (i.e. a coherent optical amplifier) shows that already SL1 is able to describe
quite accurately the output temporal profile in the presence of a  strong and rapidly varying amplification.
However, the striking power of this approach emerges when a closed-loop setup, such as the ring laser geometry, is considered.
In this case, the input field is not externally given, but determined self-consistently
from the value of the output at some previous time.
As a result, the SLn models transform into infinite dimensional delayed equations.
This is a crucial difference with the standard Galerkin truncation, which leads to a finite number of ODEs, and the higher the complexity
of the dynamics, the larger the number of modes to be retained. 
Here, the possibly high-dimensional dynamics is the result of delayed feedback, much easier to handle computationally. 
In particular, we anticipate that SL1 alone is able to
reproduce quantitatively many properties of the AB model, from the position of the second-laser threshold
over a wide range of relaxation time-scales, to the high-dimensional dynamics observed for strong pump values,
which would have otherwise required very many Fourier modes in the standard formulation of the AB model. 

Additionally, SL1 proves to be superior to the phenomenological model proposed by Vladimirov and Turaev (VT)~\cite{Vladimirov2005}.
 Starting from first-principle considerations (the AB model) and
without invoking neither the adiabatic elimination of variables nor a not-so-well defined bandwidth, SL1 provides a faster and more accurate description of semiconductor lasers.
Finally, we show that the success of our expansion in the comoving frame can be traced back to the intrinsic stability 
of the propagation along the active medium, stability which is broken only by the (delayed) feedback.

\section{Arecchi-Bonifacio model}

The starting model is the set of AB equations, whose validity is related to the accuracy of the so called slowly varying envelope approximation,
very well satisfied in the range of optical frequencies.
We refer to the formulation considered in
Ref.~\cite{Lugiato1985,Lugiato1986,DeValcarcel2003,Lugiato2010} and many other publications,
\begin{eqnarray}
\partial_z F + \partial_\tau F &=&  a(1-i\alpha)P~, \nonumber \\
\partial_\tau P &=& \gamma_\perp(DF-(1+i\tilde{\Delta})P)~, \label{oldMB}\\
\partial_\tau D &=& \gamma_\parallel \left [ 1-D - \Re(FP^*)\right ]~,\nonumber
\end{eqnarray}
where $F$ denotes the electric field, propagating along an active medium of length $L$;
$P(z,\tau)$ represents the atomic polarization, while $D(z,\tau)$ is the population inversion.
Moreover, $a$ is the pump parameter, $\gamma_\parallel$ and $\gamma_\perp$ denote the decay rate of the
population and polarization, respectively; $c$ is the speed of light and $\tilde{\Delta}$ is the detuning; 
finally, the linewidth enhancement factor $\alpha$ \cite{Henry1982} allows treating also semiconductor media.
The model evolution requires knowledge of the input field $F(0,\tau)$ for $\tau\ge 0$ and of the initial condition
$P(z,0)$, $D(z,0)$ for $0\le z\le L$.

It is worth mentioning, however, that sometimes explicit field losses can be included in the model, in the form of an additional term
$-\kappa F$ to the right-hand side of the first Eq.~(\ref{oldMB}) (this is not to be confused with the cavity losses resulting 
from the boundary conditions in closed-loop configurations).
For simplicity, we prefer not to include such a term in this study.   In section VI, we shortly comment on how it can be easily incorporated.

By rescaling and shifting the spatial variable $y = 2 z/L-1$, so that $y\in[-1,1]$, and introducing the moving frame
$t = \tau\gamma_\perp + \gamma_\perp(L - z)/c$ (the origin of $\tau$ coincides with that of $t$ at the end of the 
active medium), we obtain 
\begin{eqnarray}
\partial_y \mathcal{F}  &=& \frac{\xi}{2}\mathcal{P}, \label{MB-Feq} \\
\partial_t \mathcal{P} &=& \mathcal{D}\mathcal{F}-(1+i\tilde{\Delta})\mathcal{P}, \label{MB-Peq}\\
\partial_t \mathcal{D} &=& \gamma \left [1-\mathcal{D} - \Re(\mathcal{F}\mathcal{P^*})\right ]~, \label{MB-Deq}
\end{eqnarray}
where $\{\mathcal{F},\mathcal{P},\mathcal{D}\}$ are functions of $(y,t)$, and
$\mathcal{F}(-1, t) =F(0, t/\gamma_\perp-L/c)\equiv\mathcal{F}_a(t - T)$, where
$T = (L\gamma_\perp)/c$ is the travel time along the active medium.
Moreover, $\gamma = \gamma_\parallel/\gamma_\perp$, and $\xi=j (1-i\alpha)$, with $j = a L$.
This shows that the spatial variation of the various fields
depends on $aL$ irrespective of the specific value of the two factors, showing that 
the thin-medium limit is not well posed. The physical length $L$ contributes ``only"
to an irrelevant (in this context) time shift between the input and output signals.

\begin{figure}
\includegraphics[width=1.\linewidth]{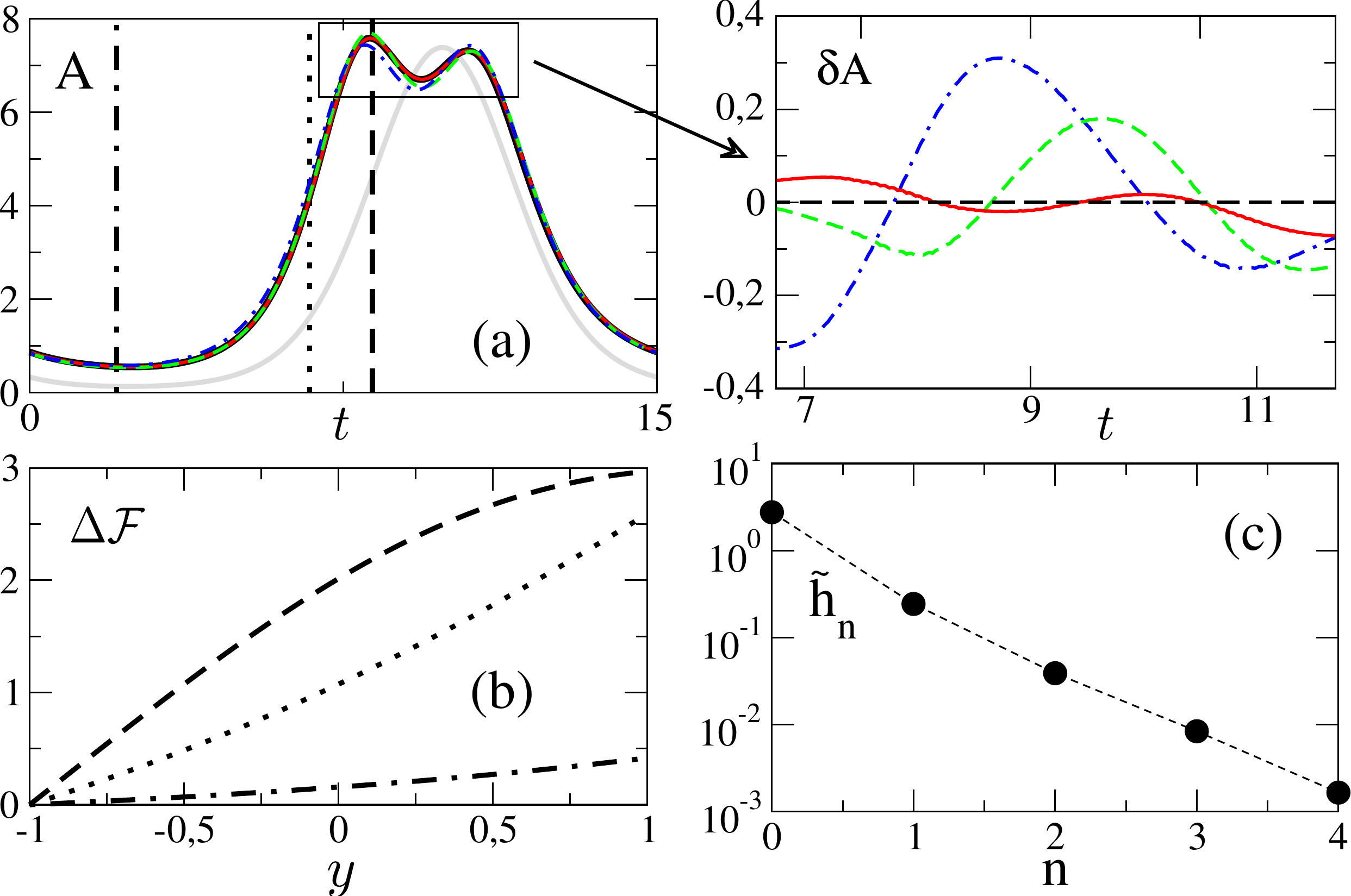}
\caption{Active medium response to the periodic signal $\mathcal{F}_a = \exp(2\sin(2\pi t/15))$ (light grey curve), for $\tilde{\Delta}=0$, $\gamma=0.1$, and
$j=4$. As $\alpha=0$, the field $\mathcal{F}$ is real with amplitude A=$|\mathcal{F}|=\mathcal{F}$. (a) Output field amplitudes $\mathcal{F}_b(t)=\mathcal{F}(1,t)$ obtained from the integration of Eqs.~(\ref{MB-Feq}-\ref{MB-Deq}) (black thick curve), SL1 (blue dot-dashed curve), SL2 (green dashed curve) and SL3 (red curve) models. In the side panel, amplitude differences $\delta$A between the AB and SLn models evaluated in the boxed region.
(b) $\Delta \mathcal{F}$ (see text) taken at the times depicted by the vertical lines in panel (a). (c) Leg\`endre spectrum.}
\label{fig1}
\end{figure}

We first consider an  active medium fed by a periodically modulated real
signal (see Fig.~\ref{fig1}); the parameters have been selected so as to have strong nonlinear effects.
The output field of the AB model is reported in panel (a), where one can
notice the qualitatively different shape of the output  with respect to the modulation. 
In the panel (b), we plot the field variation $\Delta \mathcal{F}(y,t)= \mathcal{F}(y,t)-\mathcal{F}(-1,t)$ 
along the active medium at three different times $t$. 
Notably, all profiles show a smooth, monotonic increase: 
(i) tiny for a small input field (dot-dashed);
(ii) larger for intermediate input fields (dotted curve); 
(iii) affected by saturation for yet larger amplitudes (dashed curve).

\section{Projecting the AB model on the Leg\`endre basis}

So long as this smooth dependence holds at all times, the profiles can be effectively expanded in terms of low-order polynomials. 
We have decided to use Leg\`endre polynomials (LP)~\cite{Abramowitz1965},
which, being orthogonal, are a proper basis for the projection of generic functions over a finite interval.
A similar idea has been proposed in the different context of 
stationary linear propagation along  nonuniform media~\cite{Chamanzar2006}.
Given a generic time-dependent field $\mathcal{H}(y,t)$, and denoting with
$N_n(y)$ the Leg\`endre polynomial of order $n$ defined in 
the interval 
$[-1,1]$~\footnote{At variance with the standard literature~\cite{Abramowitz1965}, where orthogonal
polynomials are defined, here we refer to orthonormal ones.},
\begin{equation}
\tilde{\mathcal{H}}_n(t) = \int_{-1}^{1}  \mathcal{H}(y,t) N_n(y)~dy
\end{equation}
identifies the instantaneous $n$-th Leg\`endre component.
We can then define the Leg\`endre spectrum
$\tilde h_n = \sqrt{ \langle \tilde{\mathcal{H}}^2_n(t) \rangle}$, where the angular brackets denote a time average. 
The spectrum of the field $\mathcal{F}$ for the periodic modulation of Fig.~\ref{fig1}(a) is reported in panel (c).
It reveals a nearly exponential decrease, which confirms the insignificance of higher order polynomials
and suggests it is worth expanding Eqs.~(\ref{MB-Feq}--\ref{MB-Deq}) into LPs. 

By representing the polarization and population as
\begin{equation}
\mathcal{P}(y,t) = \sum_{n=0} p_n(t) N_n(y)   ~,~  \mathcal{D}(y,t) = \sum_{n=0} d_n(t) N_n(y) ~,
\label{PD}
\end{equation}
the field (the integral of $\mathcal{P}$ as from Eq.~(\ref{MB-Feq})) can be expressed as
\begin{equation}
\mathcal{F}(y,t) = \mathcal{F}_a(t) + \frac{1}{2}\xi\sum_{n=0} p_n(t) \overline{N}_n(y)~,
\label{F}
\end{equation}
where $\mathcal{F}_a(t)=\mathcal{F}(-1,t)$ is the field at the beginning of the active medium, while
$\overline{N}_n(y) = \int_{-1}^{y} N_n(z)~dz~.$ 
Using  Bonnet's recursive formula~\cite{Abramowitz1965}, 
$\overline{N}_n$ can be expressed in terms of LPs, thereby obtaining an explicit expression for $\mathcal{F}(y,t)$.
Once the expansions of $\mathcal{F}$, $\mathcal{P}$, and $\mathcal{D}$ are given, we can insert them
into Eqs.~(\ref{MB-Peq},\ref{MB-Deq}) and project the resulting ODEs onto the LP basis.
The products $\mathcal{DF}$ and $\mathcal{FP}^*$ generate
terms of the type $N_m(y)N_n(y)$, which can be expressed as a linear combination 
of the $N_k(y)$ polynomials with $k\le m+n$~\cite{Neumann1878, Adams1878}.

The details of the procedure are presented in the Appendix. 
Here, we limit ourselves to
illustrate the derivation of the lowest-order model.
Since $N_0(y) =1/\sqrt{2}$ and $N_1(y) = \sqrt{3/2} y$, then
$\overline{N}_0 = N_0+ N_1/\sqrt{3}$, so that
$\mathcal{F}(y,t) = \mathcal{F}_a(t) + \frac{1}{2}\xi p_0(t) \left(y+1\right)$.
Moreover, since $N_0N_n = N_n/\sqrt{2}$, one obtains 
\begin{eqnarray}
\dot{p}_0 &=& -(1+i\tilde{\Delta})p_0 +d_0 \left(\mathcal{F}_a+
\xi p_0/2\right),\label{p0}\\
\dot{d_0} &=& \gamma \left[ 1-d_0 -\Re\left( \mathcal{F}_a p_0^* +\xi|p_0|^2/2 \right) \right].
\label{d0}
\end{eqnarray}
Finally, the field at the end of the active medium is (at any order)
\begin{equation}
\mathcal{F}_b(t) \equiv \mathcal{F}(1,t) = \mathcal{F}_a(t) + \xi p_0(t) \; ,
\label{Fb}
\end{equation}
where $\mathcal{F}_a(t)$ refers to the time $(t-T)$ in the laboratory frame.
This model is implicitly based on the assumption of a linear field-profile;
for this reason we shall refer to it as to SL1.
More in general, SLn involves $3n$ differential equations
(for $n$ real modes representing the population, and $n$ complex modes describing the polarization).

In Fig.~\ref{fig1}(a) we report the outcome of SL1-SL3 and show the corresponding difference with the AB equations
in the inset. SL1 is already able to capture the qualitative behavior of the output field, 
including the double peak. An increasingly better agreement is ensured by the higher-order models.

\section{Ring laser case}

We now turn our attention to closed-loop setups, where the input field is determined self-consistently. 
More precisely we consider ring lasers~\cite{Menegozzi1973} with unidirectional propagation.
They have been the subject of many studies:
to identify the so-called second laser threshold~\cite{Risken1968,Risken1968a,Graham1968}; 
to perform accurate stability analyses~\cite{Lugiato1985,Lugiato1986};
to derive amplitude equations~\cite{Casini1997}; 
to perform non-standard adiabatic elimination wherever appropriate~\cite{DeValcarcel2003,Perego2020}, or to study temporal localized states \cite{Vladimirov2005,Schelte2018}. 
The abundance of results make the ring laser an optimal testing ground for our approach.

In a ring laser, $\mathcal{F}_a(t) = R \mathcal{F}_b(t-T_r)$, where $R$ is the reflectivity of the mirror(s), while
$T_r = T + T_f$ is the round-trip time, $T_f$ being the free propagation time from the end of the active medium 
back to the origin. 
A similar model was proposed by Milonni et al.~\cite{Milonni1987}, 
who derived their equations under the thin-medium approximation: an ill-posed assumption since we have seen that the
spatial dependence cannot be controlled by the physical length alone, once the pump parameter $\xi$ is given.

The ring condition has turned the original set of ordinary equations into a delayed equation, known to be
infinite dimensional~\cite{Hale1993,Erneux2009,Yanchuk2017,Hart2019a}.
This property is crucial since it allows SL1 reproducing the richness of the original AB model, in spite of its low
computational complexity. 

The model structure is better appreciated by
eliminating $\mathcal{F}_a$ from Eqs.~(\ref{p0},\ref{d0}) with the help of Eq.~(\ref{Fb}).
As a result, we obtain the first order, ring laser (RL1) model
\begin{eqnarray}
&& \mathcal{F}_b(t) = R \mathcal{F}_b(t-T_r) + \xi p_0(t),\\
\dot{p}_0 &=& -(1+i\tilde{\Delta})p_0 +d_0 \big(\mathcal{F}_b-
\xi p_0/2\big),\\
\dot{d_0} &=& \gamma \left[ 1-d_0 -\Re\left( \mathcal{F}_b p_0^* -\xi|p_0|^2/2 \right) \right].
\end{eqnarray}

\begin{figure}
\includegraphics[width= 0.8\linewidth]{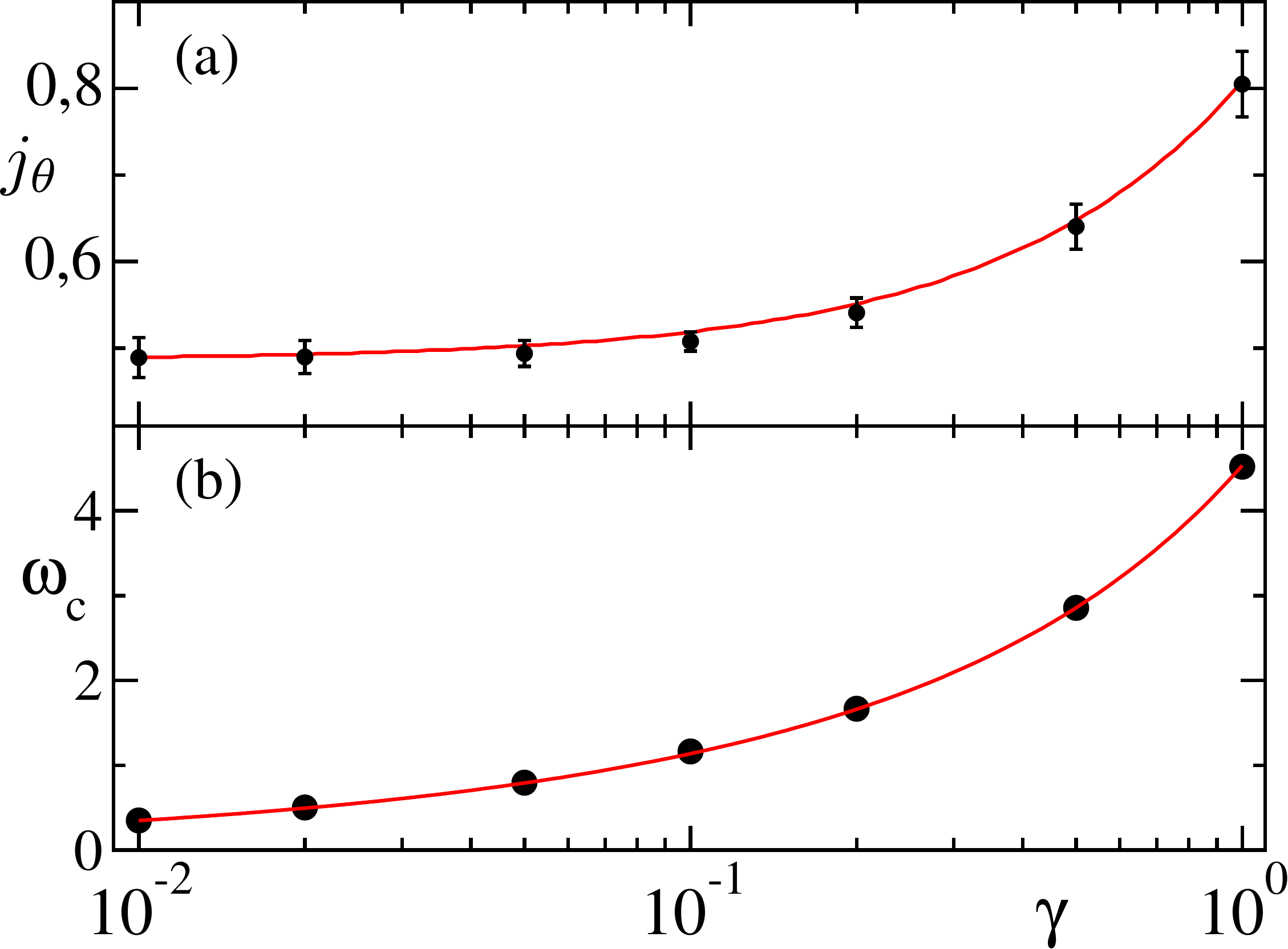}
\caption{Second laser threshold for a ring laser with $\alpha=0$. (a) Critical pump value and (b) corresponding frequency.
Data refer to $R=0.95$ in the limit of a long delay (see text). Solid curve: solution of the analytic
expressions of \cite{Lugiato1986}; dots: numerical solution of the RL1 model.}
\label{fig2}
\end{figure}

We now test the ring-laser dynamics by determining the second laser threshold, where the stationary state destabilizes. 
An analytic characteristic equation is available for the linearized AB system when $\alpha=0$~\cite{Lugiato1986}.
The numerical solution of such equation is plotted in Fig.~\ref{fig2} for $R=0.95$.
The threshold values $j_{\theta}$, and the corresponding frequency of the leading unstable mode are shown as solid 
curves in Fig.~\ref{fig2}(a,b), for a broad range of $\gamma$ values. The full circles in the same figure have been obtained by numerically integrating the RL1 model.  We have used a long delay ($T_r=200-800$), in order to better resolve the critical point thanks to the high modal density \cite{Hale1993,Erneux2009,Yanchuk2017}. The error bars are due to the difficulty of discriminating whether perturbations do grow or converge to zero in the vicinity of the bifurcation. As seen in the figure, the agreement is excellent already using the lowest order of approximation.

Then, we have considered the semiconductor setup  ($\alpha=5$) to test a different system 
and to compare with a pre-existing delayed model proposed to characterize 
small-$\gamma$ devices, where the polarization has been adiabatically eliminated: the VT 
model~ \cite{Vladimirov2005}.
Using the notations of this paper, the VT model can be written as
$ \dot{\mathcal{F}}_b = - \mathcal{F}_b + R \mathrm{e}^{\xi \mathcal{G}}\mathcal{F}_b(t-T_r)$ plus
$\dot{\mathcal{G}} = \bar{\gamma} [1- \mathcal{G} + (1- \mathrm{e}^{2j\mathcal{G}} ) |\mathcal{F}_b(t-T_r)|^2]$,
where $\mathcal{G}$ is the spatial integral of the population inversion and $\bar{\gamma}$ is a phenomenological parameter
playing a role similar to $\gamma$.
For $R=0.95$, $T_r=200$, and  $\gamma = 10^{-3}$, both the AB and RL1 models reveal that stationary states lose stability
above a critical amplification $j_\theta$, approximately equal to 0.25 and 0.17, respectively.
On the other hand, the integration of the VT model does not reveal any destabilization up to $j = 5$ for a range of
$\bar{\gamma}$ values from $10^{-2}$ to $10^{-4}$. We are led to conclude that 
our RL1 model is more accurate than VT, at least in the considered setup~\footnote{One should remember that VT model was derived in the additional presence of a saturable absorber.}.

\begin{figure}
\includegraphics[width= 1.2 \linewidth]{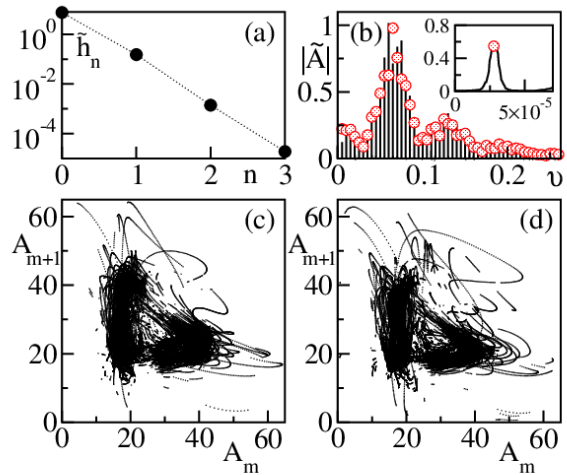}
\caption{Comparison of RL1 and AB models for a semiconductor medium ($\alpha=5$) well beyond the second threshold. Here, $j=2.5$, $T_r=200$ and  $\gamma = 10^{-3}$. (a) Leg\`endre spectrum from the AB integration. (b) Modulus of the power spectrum of the field amplitude A =$|\mathcal{F}|$ (field is now complex): AB (solid black lines) versus SL1 (red dots); inset: zoom of the first peak. Poincar\'e sections of successive maxima of A for AB (c) and RL1 (d) model.}
\label{fig3}
\end{figure}

Finally, we have made a more stringent test, simulating the laser significantly above threshold, where the dynamics is irregular. In Fig.~\ref{fig3}(a), we report the Leg\`endre spectrum obtained by
integrating the AB equations. Analogously to the open loop setup, we observe a clean fast exponential decay:
a strong hint that our approach is going to work. In panel (b) we 
superpose the Fourier amplitude of the field dynamics obtained from the AB equation (solid line) with
the peaks of the spectrum obtained from the RL1 model: the agreement is remarkable.
Since the Fourier spectrum does not contain information on the phases, we have also constructed a Poincar\'e section
from the maxima of the field amplitude. The results from the AB model are presented in panel (c) to be compared with the
outcome of the RL1 model, presented in panel (d). This comparison confirms the validity of the approximate
model.

\section{Modal expansion and generalized synchronization}

We lastly discuss the origin of the success of the modal expansion, starting from 
the response of an active medium to a generic time-dependent field $\mathcal{F}_a$. 
According to Eqs.~(\ref{MB-Feq}-\ref{MB-Deq}), the medium can be seen as
a series of identical slices along the $y$ direction, each slice being modulated
by a field, made of two components
$\mathcal{F}(y,t) =\mathcal{F}_a(t) + \Delta \mathcal{F}(y,t)$, where 
$\Delta \mathcal{F}$ is the integral of the polarization for $x\le y$.
The unidirectionality of the coupling implies that the overall dynamics can be assessed by separately looking at the single slices for fixed $y$.
For a fixed slice (fixed $y$) and given $\mathcal{F}(y,t)$, the polarization $\mathcal{P}(y,t)$ and the population $\mathcal{D}(y,t)$  follow a linear dynamics (\ref{MB-Peq},\ref{MB-Deq}), and can therefore be treated analytically.
In particular, the positive-definite observable
$\mathcal{L} = \gamma |\mathcal{P}|^2 + \mathcal{D}^2$ is a proper Lyapunov function \cite{Shilnikov2001,Wiggins2003} for the homogeneous part of the equations (\ref{MB-Peq},\ref{MB-Deq}) 
\begin{equation}
\dot{\mathcal{L}} = -2 \gamma (|\mathcal{P}|^2 +\mathcal{D}^2) \le -2 \gamma_m \mathcal{L} \; ,
\label{eq:Lyapf}
\end{equation}
where $\gamma_m = \min\{\gamma,1\}$. The inequality (\ref{eq:Lyapf}) can be proven by direct substitution. 
The derivative of $\mathcal{L}$ is negative and uniformly bounded, 
indicating an exponential decay to zero.

Hence, dynamical degrees corresponding to the variables $\mathcal{P}$ and $\mathcal{D}$ do not contribute to the active dynamical degrees of the spatially-extended active medium.
More precisely, the Lyapunov function (\ref{eq:Lyapf}) implies that, at any given slice $y$, the polarization $\mathcal{P}$ and population $\mathcal{D}$ are synchronized in generalized sense    \cite{Abarbanel1996a,Kocarev1996,Pikovsky2001,Uchida2003} to a given field $\mathcal{F}(y,t)$. 
The property of the generalized synchronization implies that the polarization and population variables are uniquely determined by the field variable, and no additional active degrees of freedom emerge.
%On the other hand, the long-range (integral) nature of $\Delta \mathcal{F}(y)$ instantaneously smooths the field profile,
%again consistently with the absence of oscillations in the comoving frame.
As a result, the instabilities arising in closed-loop configurations are entirely due to the delayed feedback.
Herein lies the superiority of our approach: convective instabilities arising in the original formulation
(see Eq.~(\ref{oldMB})) are converted into a delayed-induced instability, accompanied by a spatial stability.
This is not a surprise in the long delay limit, as it is well known that delay may induce ``convective" instabilities \cite{Giacomelli1996},
but it is true also in the short delay limit, when the ring condition reduces to an ODE,
$\dot{\mathcal{F}}_b = (-(1-R)\mathcal{F}_b + \xi p_0)/(RT_r)$. Indeed, this equation, accompanied by Eqs.~(\ref{p0},\ref{d0})
coincides with the Lorenz-Haken model \cite{Haken1975,Lorenz1963} under the additional approximation of negligible
$\xi$-terms in Eqs.~(\ref{p0},\ref{d0}) (typically valid in the so-called uniform field limit).

\section{Summary}

In this paper, we have introduced an effective approach which simplifies the treatment of optical active media by eliminating 
the spatial dependence. The method proves to be very accurate and fast to simulate, while retaining the richness of the full AB model. 

We have neglected exlicit field losses, but they can be easily accounted for by including a linear term in Eq.~(\ref{MB-Feq})
and thereby modifying the expansion (\ref{F}). It is important, moreover to stress that while 
the length of the medium is a meaningless concept in the absence of propagation losses,
it becomes important when they are included.

We have also assumed a constant (in space) pump $a$, but one can easily include non uniformities so long
as the pump profile can be effectively expanded into Leg\`endre polynomials.
The additional complexity would be equivalent to that of the quadratic nonlinearities already present in the original equations.

An interesting question concerns the number of modes to be accounted for. We have seen that already the
simplest model is able to reproduce the expected dynamics in a wide range of physical conditions.
This includes semiconductor ring lasers, where it proves superior to the VT model (it will be, nevertheless, worth
including a saturable absorber in our model to perform a more compelling test).
We envisage that higher-order  polynomials might be required in the presence of a large pump
in bad cavity limit~\cite{BCL}, because of the strong amplification across the active medium.

Finally, bi-directional propagation is perhaps the most interesting challenge. The elimination of spatial propagation in
ultra-thin media proposed in Ref.~\cite{Lugiato2010} seems to be a useful starting point.

\begin{acknowledgments}
	GG and AP are indebted to F.T. Arecchi for past illuminating discussions. 
	SY was supported by the German Research Foundation DFG, Project
 411803875. 
\end{acknowledgments}

\section*{Appendix: Projecting the Arecchi-Bonifacio equations on a Leg\`endre basis}
Here, we introduce the relationships required to derive the evolution equations at any prescribed order.
Most of them are known formulas, herewith recalled to help the reader. 

The third-order equations (model SL3) are finally presented to exemplify the outcome of the procedure.

\subsection{Legendre polynomial basis definition}
\noindent 
In the scientific litereture, the Leg\`endre polynomials $Q_n(y)$ are defined as an orthogonal basis on the interval
$[-1,1]$ (see e.g. \cite{Abramowitz1965}).
Here we prefer to refer to the orthonormal polynomials $N_n(y)$,
\begin{equation}
\int_{-1}^{1} N_n(y)N_m(y)~dy = \delta_{nm}~.
\end{equation} 
The two sets of polynomials differ only by a scaling factor, 
\begin{equation}
N_n(y) = \sqrt{\frac{2n+1}{2}} ~ Q_n(y) ~,
\end{equation}
In particular, 
\begin{eqnarray}
N_0(y) &=& \frac{1}{\sqrt{2}},  \\ 
N_1(y) &=& \sqrt{\frac{3}{2}}~ y, \nonumber \\ 
N_2(y) &=& \frac{1}{2}\sqrt{\frac{5}{2}}~(3y^2 -1), \nonumber \\ N_3(y) &=& \frac{1}{2}\sqrt{\frac{7}{2}}(5 y^3-3y),  ~\ldots \nonumber
\end{eqnarray}

\subsection{Integral of Legendre polynomials}
\noindent
One of the key expressions required to expand the AB model concerns the spatial integration of the polynomials,
\begin{equation}
\overline{N}_n(y) = \int_{-1}^{y} N_n(z)~dz = \sqrt{\frac{2n+1}{2}} \overline{Q}_n(y) \; .
\end{equation}

It is known that the $Q_n$ polynomials satisfy the differential equation~\cite{Abramowitz1965}
\begin{equation}
\frac{d}{dy} \left[ (1-y^2) Q_n(y) \right] + n(n+1)Q_n(y) = 0 \; .
\end{equation}
Hence,
\begin{equation}
\overline{Q}_n(y) = \frac{y^2-1}{n(n+1)} \frac{dQ_n(y)}{dy} ~.\,
\end{equation}
Next, recalling that~\cite{Abramowitz1965}
\begin{equation}
\frac{y^2-1}{n} \frac{dQ_n(y)}{dy} = yQ_n(y)-Q_{n-1}(y)  \; ,
\end{equation}
we obtain
\begin{equation}
\label{eq:qover}
\overline{Q}_n(y) = \frac{yQ_n(y)-Q_{n-1}(y)}{n+1}.
\end{equation}
Bonnet's relationship~\cite{Abramowitz1965} (valid for $n>0$), allows removing the explicit $y$ dependence,
\begin{equation}
\label{eq:bonnet}
yQ_n(y) = \frac{n}{2n+1}Q_{n-1}(y)+\frac{n+1}{2n+1}Q_{n+1}(y) ~.
\end{equation}
In fact, by inserting Eq.~(\ref{eq:bonnet}) into Eq.~(\ref{eq:qover}), we obtain
\begin{equation}
\overline{Q}_n(y) = \frac{Q_{n+1} - Q_{n-1}}{2n+1}.
\end{equation}
Finally, referring to the orthonormal polynomials
\begin{equation}
\label{eq:iter}
	\overline{N}_n(y) = \frac{1}{\sqrt{2n+1}} \left[\frac{N_{n+1}(y)}{\sqrt{2n+3}} -
	\frac{N_{n-1}(y)}{\sqrt{2n-1}} \right]
.
\end{equation}

The first integral ($n=0$) must be computed directly
\begin{equation}
\overline{N}_0(y) = \frac{1}{\sqrt{3}}N_1(y) + N_0(y) \; .
\end{equation}
The other integrals are thereby recursively obtained from Eq.~(\ref{eq:iter}), starting from 

\begin{equation}
\overline{N}_1(y) = \frac{1}{2}\sqrt{\frac{3}{2}} (y^2- 1) = \frac{1}{\sqrt{15}}N_2(y) -
\frac{1}{\sqrt{3}}N_0(y)  ~.
\end{equation}

\subsection{Projection of nonlinear terms}

\noindent In order to project the nonlinear terms present in the polarization and population equations, it is necessary to express the 
product of two Leg\`endre polynomials in terms of Leg\`endre polynomials themselves.
A general formula was given in 1878 independently by F. E. Neumann \cite{Neumann1878} and J.C. Adams \cite{Adams1878}, 
and proved later e.g. by W.A. Salam \cite{Al-salam1956},
\begin{widetext}
\begin{equation}
Q_p(x) Q_q(x) = 
\sum_{r=0}^q \frac{ A_r A_{p-r} A_{q-r}}{A_{p+q-r}} \frac{2p+2q-4r+1}{2p+2q-2r+1} Q_{p+q-2r}(x)~,
\end{equation}
with 
\begin{equation}
p \geq q \quad A_r = \frac{(\frac{1}{2})_r}{r!}, \quad 
(a)_r =a(a+1)..(a+r-1), \quad  \textrm{and} \quad (a)_0=1~.\nonumber
\end{equation}
By normalizing, we obtain the required expression,
\begin{equation}
	N_p(x) N_q(x) = \sqrt{\frac{(2p+1)(2q+1)}{2\big(2(p+q-2r)+1\big)}}\times
	\sum_{r=0}^q \frac{ A_r A_{p-r} A_{q-r}}{A_{p+q-r}} \frac{2p+2q-4r+1}{2p+2q-2r+1} 
	N_{p+q-2r}(x)~.
\end{equation}
\end{widetext}

\subsection{The evolution equations}
By using the general relationships given in the previous section, it is possible to obtain approximate models, by
truncating the hierarchy of equations at the desired order.
Here we derive SL3.
For the sake of completeness and clarity, the amplification factor is expressed explicitly in terms of
the pump $j_0$ and the Henry's $\alpha$ factor, $\xi=j_0(1-i\alpha)$. The equations are,
\begin{widetext}
\[
\dot{p}_{0}=-\left(1+i\tilde \Delta\right)p_{0}+\mathcal{F}_{a}d_{0}+\frac{j_{0}}{2}\left(1-i\alpha\right)\left(d_{0}p_{0}-\frac{1}{3}d_{0}p_{1}+\frac{1}{3}d_{1}p_{0}{\color{black}-\frac{1}{15}d_{1}p_{2}+\frac{1}{15}d_{2}p_{1}}\right),
\]
\[
\dot{p}_{1}=-\left(1+i\tilde \Delta\right)p_{1}+\mathcal{F}_{a}d_{1}+\frac{j_{0}}{2}\left(1-i\alpha\right)\left(d_{0}p_{0}+d_{1}p_{0}-\frac{1}{5}d_{1}p_{1}{\color{black}-\frac{1}{5}d_{0}p_{2}+\frac{2}{5}d_{2}p_{0}-\frac{1}{35}d_{2}p_{2}}\right),
\]

\[
{\color{black}\dot{p}_{2}=-(1+i\tilde \Delta)p_{2}+\mathcal{F}_{a}d_{2}+\frac{j_{0}}{2}\left(1-i\alpha\right)\left(\frac{1}{3}d_{0}p_{1}+\frac{2}{3}d_{1}p_{0}+d_{2}p_{0}-\frac{1}{21}d_{1}p_{2}-\frac{5}{21}d_{2}p_{1}\right)},
\]
\[
\dot{d}_{0}=\gamma\left[ 1-d_{0}-\Re\left(\mathcal{F}_{a}\overline{p_{0}}\right)-\frac{j_{0}}{2}\left|p_{0}\right|^{2}-\frac{\alpha j_{0}}{3}\Im\left(p_{0}\overline{p_{1}}\right){\color{black}-\frac{\alpha j_{0}}{15}\Im\left(p_{1}\overline{p_{2}}\right)} \right],
\]
\[
\dot{d}_{1}=\gamma\left[-d_{1}-\Re\left(\mathcal{F}_{a}\overline{p_{1}}\right)+\frac{j_{0}}{2}\left(-\left|p_{0}\right|^{2}-\Re\left(\left(1+i\alpha\right)p_{1}\overline{p_{0}}\right)+\frac{1}{5}\left|p_{1}\right|^{2}{\color{black}+\frac{1}{35}\left|p_{2}\right|^{2}-\frac{1}{5}\Re\left(\left(1-3i\alpha\right)p_{0}\overline{p_{2}}\right)}\right) \right],
\]
\[
{\color{black}\dot{d}_{2}=\gamma\left[-d_{2}-\Re\left(\mathcal{F}_{a}\overline{p_{2}}\right)+\frac{j_{0}}{2}\left(-\frac{1}{3}\Re\left(\left(3-i\alpha\right)p_{0}\overline{p_{1}}\right)-\Re\left(\left(1-i\alpha\right)p_{0}\overline{p_{2}}\right)+\frac{2}{21}\Re\left(\left(3-2i\alpha\right)p_{1}\overline{p_{2}}\right)\right)\right]}.
\]
The instantaneous field profile is 
\[
\mathcal{F}(y)=\mathcal{F}_{a}+\frac{j_{0}}{2}\left(1-i\alpha\right)\left(p_{0}\left(1+y\right)+\frac{p_{1}}{2}\left(y^{2}-1\right){\color{black}+\frac{p_{2}}{2}\left(y^{3}-y\right)}\right)  \, .
\]
\end{widetext}

Typically, one is interested in the field amplitude at the end of the active medium,
\[
\mathcal{F}_b= \mathcal{F}(1) = \mathcal{F}_{a}+j_{0}\left(1-i\alpha\right)p_{0} \; ,
\]
which depends only on the input field and the zeroth polarization mode. This is true at any order, since all Leg\`endre polynomial
have zero average, except for $N_0(y)$.

By omitting the 
%blue-colored 
terms (and related equations) containing the variables $p_2$ and $d_2$, one obtains SL2, the model of order 2.
SL1, defined by Eq.~(\ref{F}), can be obtained from the above equations by omitting 
all terms containing $p_1,p_2,d_1,d_2$ and their corresponding equations. 

Finally, note that the truncated models possess the following general form:
\begin{eqnarray*}
	\dot{\boldsymbol{p}}&=&-\left(1+i\tilde \Delta\right)\boldsymbol{p}+\mathcal{F}_{a}\boldsymbol{d}+j_{0}\left(1-i\alpha\right)B_{1}(\boldsymbol{d},\boldsymbol{p}), \\
	\dot{\boldsymbol{d}} &=&\gamma\left[e_{0}-\boldsymbol{d}-\Re\left(\mathcal{F}_{a}\overline{\boldsymbol{p}}\right)+j_{0}B_{2}(\boldsymbol{p},\overline{\boldsymbol{p}})\right],
\end{eqnarray*}
where $\boldsymbol{p}=(p_1,\dots,p_M)^T$,
$\boldsymbol{d}=(d_1,\dots,d_M)^T$, 
$e_0=(1,0,\dots,0)^T$, while $B_1$ and $B_2$ are bilinear forms, and $M$ is the truncation order.

\bibliographystyle{apsrev4-1}

%\bibliography{/home/yanchuk/Documents/bibfiles/library}

%

\end{document}